\documentclass[a4paper,11pt]{article}
\usepackage{graphicx,amssymb,amstext,amsmath}
\usepackage{color}
\usepackage{floatflt}

\definecolor{cbl}{rgb}{0,0,1}                

\newcommand{\bc}{\begin{center}}
\newcommand{\ec}{\end{center}}
\def\ba#1{\begin{array}{#1}\displaystyle}
\newcommand{\ea}{\end{array}}

\newcommand{\beq}{\begin{equation}}
\newcommand{\eeq}{\end{equation}}
\newcommand{\beqa}{\begin{eqnarray}}
\newcommand{\eeqa}{\end{eqnarray}}

\newcommand{\n}{\nonumber\\}
\newcommand{\bi}{\begin{itemize}}
\newcommand{\ei}{\end{itemize}}

\def\lt#1{\left#1}
\def\rt#1{\right#1}
\def\t#1{\tilde{#1}}

\def\b#1{\bar{#1}}
\def\frc#1#2{\frac{#1}{#2}}

\newcommand{\p}{\partial}

\newcommand{\bra}{\langle}
\newcommand{\ket}{\rangle}

\newcommand{\R}{{\mathbb{R}}}
\newcommand{\C}{{\mathbb{C}}}

\newcommand{\dd}{\mathrm{d}}

\newcommand{\Or}{{\cal O}}

\newcommand{\hash}{\#}

\newcommand{\real}{{\mathsf R}}

\begin{document}
\begin{titlepage}
\vspace{0.2cm}
\begin{center}

{\Large{\bf{ Irreversibility of the renormalization group flow\\[0.2cm] in non-unitary quantum field theory}}}

\vspace{0.8cm} 
{\large \text{Olalla A. Castro-Alvaredo${}^{\heartsuit}$, Benjamin Doyon{\LARGE${}^{\star}$} and Francesco Ravanini{$\,{}^{\spadesuit}$}}}

\vspace{0.8cm}
{\small{{\small ${}^{\heartsuit}$} Department of Mathematics, City, University of London, 10 Northampton Square EC1V 0HB, UK }\\
\vspace{0.2cm}
{{\Large ${}^{\star}$} Department of Mathematics, King's College London, Strand, London WC2R 2LS, UK}\\
\vspace{0.2cm}
{{ ${}^{\spadesuit}$} Dipartimento di Fisica e Astronomia, Universit\`a di Bologna, Via Irnerio 46, I-40126 Bologna, Italy and  INFN, Sezione di Bologna, Via Irnerio 46, I-40126 Bologna, Italy}}

\end{center}

\vspace{1cm}
We show irreversibility of the renormalization group flow in non-unitary but ${\cal PT}$-invariant quantum field theory in two space-time dimensions. In addition to unbroken $\mathcal{PT}$-symmetry and a positive energy spectrum, we assume standard properties of quantum field theory including a local energy-momentum tensor and relativistic invariance. This generalizes Zamolodchikov's $c$-theorem to ${\cal PT}$-symmetric hamiltonians. Our proof follows closely Zamolodchikov's arguments. We show that a function $c_{\mathrm{eff}}(s)$ of the renormalization group parameter $s$ exists which is non-negative and monotonically decreasing along renormalization group flows. Its value at a critical point is the ``effective central charge" entering the specific free energy.  At least in rational models, this equals $c_{\mathrm{eff}}=c-24\Delta$, where $c$ is the central charge and $\Delta$ is the lowest primary field dimension in the conformal field theory which describes the critical point.
\medskip

\vfill

\noindent 
{\small ${}^{\heartsuit}$ o.castro-alvaredo@city.ac.uk\\
{\Large{${}^{\Large\star}$}} benjamin.doyon@kcl.ac.uk\\ 
${}^{\spadesuit}$ francesco.ravanini@unibo.it
\hfill \today}

\end{titlepage}

\hfill{\it Dedicated to John Cardy on the occasion of his 70th birthday}

\section{Introduction}
The theoretical understanding of quantum critical phenomena is one of the cornerstones of modern theoretical physics. It has been known for some time that such phenomena, both classical and quantum, are characterized by universal laws described by conformal field theory (CFT) \cite{DMS}.  Conformal invariance is constraining enough as to fully determine the space-time dependence of two- and three-point functions of local fields, demonstrating that they must exhibit power-law decay, a defining feature of criticality. Conformal field theory is particularly powerful in two dimensions \cite{BPZ}, where it is associated with the presence of infinite-dimensional symmetry algebras. Beyond criticality, universal behaviours subsist  at large observation scales when correlation lengths are much larger than microscopic distances. These are described by non-conformal quantum field theory (QFT). One of the deepest ideas in this context is that of the renormalization group (RG). RG flows describe how, in parameter space, a QFT model varies with the scale, connecting an ultraviolet (small length scale) to an infrared (large length large scale) fixed-point CFT. These give  rise to flows between CFT models, and one might wonder what structure such RG flows may take.


In the eighties Alexander B. Zamolodchikov \cite{zamc} showed that the space of {\it unitary} CFT is  partially ordered by unitary RG flows: an RG flow associated to a non-conformal unitary QFT may exist from an ultraviolet (UV) to an infrared (IR) fixed point only if the central charges of the corresponding CFTs are ordered,
\beq\label{ineq}
	c_{\rm UV} > c_{\rm IR}.
\eeq
This implies irreversibility of (nontrivial) RG flows.
Specifically, he constructed for any 1+1 dimensional unitary QFT, a function $c(s)$ (called $c$-function, or scaling function) of the renormalization group parameter $s=2\log(mr)$, where $m$ is a characteristic mass scale and $r$ is the observation length scale, with the following properties: it is non-negative for all $s$, it is monotonically strictly decreasing along the RG flow (from the UV to the IR) and it takes constant values  at critical points, given by the central charges of the corresponding CFTs. Therefore, the function ``flows" between its ultraviolet (or high-energy) $c_{\rm UV}=\lim_{s\rightarrow -\infty}c(s)$ and its infrared (or low-energy) $c_{\rm IR}=\lim_{s\rightarrow \infty}c(s)$ values, which are the central charges of the two CFTs that are found at high and low energies. This implies the strict inequality \eqref{ineq}.

This statement is known as the $c$-theorem and it constitutes one of the most fundamental results for 1+1 dimensional QFT. Several alternative proofs of the $c$-theorem exist which, for instance, employ finite-size field-theory methods \cite{cth2} and holographic arguments for the entanglement entropy \cite{Casinic}.  
The function $c(s)$ may be interpreted as counting degrees of freedom at a given energy scale\footnote{This interpretation is particularly well illustrated for some classes of integrable models \cite{roaming, CastroAlvaredo:2000ag} where the $c$-function visits the vicinity of multiple critical points between its UV and its IR values (the function considered in \cite{roaming} is a different $c$-function than that defined in \cite{zamc}, which however takes the same values at critical points).}. It also has the interpretation as an off-critical Casimir energy, since the Casimir energy of CFT on a cylinder is proportional to the central charge $c$ \cite{BCN,Affleck}, and it characterizes the logarithmic growth of entanglement in one-dimensional quantum critical systems  \cite{HolzheyLW94, Latorre1, Calabrese:2004eu}. The key role played by the central charge $c$  in the description of critical phenomena has been nicely reviewed in John Cardy's Boltzmann Medal lecture \cite{ubi}.

In higher dimensions, similar concepts arise. Starting with John Cardy's work \cite{higherd1}, the existence in four dimensional theories of an $a$-function with similar properties to Zamolodchikov's $c$-function has been recently proven  \cite{atheorem}.
The existence of a similar monotonic function of the RG flow in three dimensions, known as an $F$-function, was conjectured in \cite{Ftheorem} and later proven \cite{Ftheorem2} by using concepts of holography. Similar holographic arguments were also employed in \cite{Casinia} to provide an alternative proof of the $a$-theorem. Entropic proofs  exploit the properties of the bi-partite entanglement entropy in unitary QFT (e.g.~subadditivity), emphasizing its interpretation as counting the number of degrees of freedom.

In \cite{zamc} Zamolodchikov also provided a precise construction procedure for the function $c(s)$ which employs the following correlators:
\begin{eqnarray}
&&F(z\bar{z}):= z^4 \langle  T(z, \bar{z}) T(0,0) \rangle=\bar{z}^4 \langle  \bar{T}(z, \bar{z}) \bar{T}(0,0) \rangle, \label{f}\\
&&G(z\bar{z}):=z^3 \bar{z}\langle T(z, \bar{z})\Theta(0,0) \rangle=\bar{z}^3 {z}\langle \bar{T}(z, \bar{z})\Theta(0,0) \rangle, \label{gg}\\
&& H(z\bar{z}):= z^2 \bar{z}^2 \langle  \Theta (z, \bar{z})\Theta(0,0) \rangle,\label{hhh}
\label{fgh}
\end{eqnarray}
\noindent in  terms of the usual complex coordinates $z=x+iy, \bar{z}=x-iy$ (where $y$ is imaginary time). The operators above are nothing but the various components of the energy-momentum tensor $T_\mu^\nu$ in these variables, namely:
$T(z,\bar{z}):= T_{zz}(z,\bar{z})$,  
$ \bar{T}(z,\bar{z}):=T_{\bar{z}\bar{z}}(z,\bar{z})$ and the trace
$T^\mu_\mu(z,\b z) = \Theta(z,\bar{z})=4 T_{z\bar{z}}(z,\bar{z}) = 4T_{\bar{z}z}(z,\bar{z})$.  Employing conservation of the energy-momentum tensor $\bar{\partial}T(z,\bar{z})+\frac{1}{4}\partial \Theta(z,\bar{z})=0$  and $\partial\bar{T}(z,\bar{z})+\frac{1}{4}\bar\partial\Theta(z,\bar{z})=0$, where $\partial:=\frac{\partial}{\partial z}$ and $\bar{\partial}:=\frac{\partial}{\partial \bar{z}}$ it then follows that the function
\beq
	c(s)=4\pi^2\left(2 F(r)-G(r)-\frac{3}{8}H(r)\right),
	\label{tada}
\eeq
with $r=z\bar{z}$ and $s=2\log(mr)$, satisfies the equation:
\beq
\frac{dc}{ds}=-3\pi^2H(r).
 \eeq 
Since $\Theta$ is a hermitian operator, the right-hand side is non-negative by reflection positivity of unitary theories. Thus $c(s)$ is monotonically decreasing. It is well known that $H(r)$ vanishes at critical points (where the trace of the stress-energy tensor is vanishing). In addition, it is also known that at critical points, $\Theta$ vanishes and
 \beq 
4\pi^2 \langle T(0) T(z) \rangle=\frac{c}{2z^4},
\label{7}
 \eeq 
where $c$ is the central charge, and so $4\pi^2 F(z\bar{z})=\frac{c}{2}$ at critical points. Note that the unusual $4\pi^2$ factor is due to the normalization of the energy-momentum tensor chosen above, which differs from the standard normalization $T(z,\bar{z})=-2\pi T_{zz}$ (see e.g.~\cite{DMS}).
This means that $c(s)$ as defined above satisfies all three requirements of Zamolodchikov's $c$-theorem. 

 As powerful as this result is,  one may wonder if a proof of existence of such an RG scaling function may be found under less stringent conditions. There is a wide class of QFTs in 1+1 dimensions which are non-unitary: in their usual CFT description in terms of Virasoro representations, negative-norm states exist. Yet in many cases the spectrum is real and bounded from below, and such QFTs describe {\em bona fide}, near-critical points of local quantum models with non-hermitian but positive-spectrum hamiltonian acting on a proper Hilbert space. A good example is the famous Lee-Yang model of CFT which describes the Lee-Yang edge singularity \cite{Fisher,LYCardy}. It has an integrable massive QFT counterpart which leads to a perfectly reasonable QFT \cite{Cardymuss, TCSA, Z} (a good discussion of non-unitarity and scattering theory in the context of massive integrable QFTs can be found in \cite{TW,TW2}).
Several manifestly non-hermitian discrete implementations of the Lee-Yang model exist, such as the quantum spin chain studied in \cite{gehlen1,meandreas}, which possesses an Ising critical point and a Lee-Yang critical line. A non-unitary QFT describing the near-critical region of this spin chain, including the Ising $\mapsto$ Lee-Yang flow, was studied in \cite{fonseca}.

May we also define a $c$-function for these theories? If so, how can we establish that it is indeed a $c$-function without appealing to unitarity of the QFT?
 
 Non-unitary 1+1 dimensional CFTs abound (e.g. there are infinitely many examples within the minimal models of CFT) and much is known about their properties.  A feature that has emerged in various contexts is that for such theories, the role of the central charge $c$ is taken up by an effective central charge $c_{\mathrm{eff}}:=c-24\Delta$ where $\Delta$ is the smallest conformal dimension of the spectrum of local fields. For unitary theories $\Delta=0$, corresponding to the identity field but for non-unitary models $\Delta$ and $c$ may both be negative in such a way as to give $c_{\mathrm{eff}}\geq 0$. For instance, the ground state energy of a non-unitary CFT on the cylinder is proportional to $c_{\mathrm{eff}}$ \cite{ceff} whereas for unitary theories it is proportional to $c$ \cite{BCN,Affleck}. Similarly, the von Neumann entanglement entropy of connected sub-system in a 1+1 dimensional CFT diverges logarithmically with the size of the sub-system with a coefficient which is proportional to $c$ in unitary theories \cite{HolzheyLW94, Latorre1, Calabrese:2004eu} and to $c_{\mathrm{eff}}$ in non-unitary ones \cite{BCDLR,BR,Saleur}.  These results all highlight the important role played by $c_{\mathrm{eff}}$ at criticality for non-unitary CFTs.
 
 Furthermore, there are many known examples of RG flows between such CFTs with the property 
  \beq 
 (c_{\mathrm{eff}})_{\rm UV}> (c_{\mathrm{eff}})_{\rm IR},
 \label{order}
 \eeq 
 hence generalising (\ref{ineq}) for non-unitary models. 
 {\begin{floatingfigure}[h!]{7cm} 
 \begin{center} 
 \includegraphics[width=7cm]{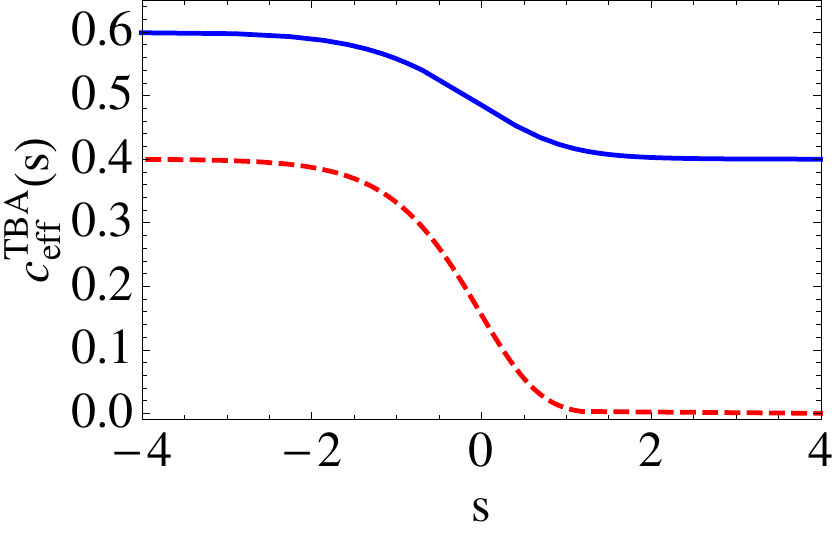} 
 \end{center} 
 \caption{Typical RG flows from and to the Lee-Yang model $\mathcal{M}_{2,5}$ with $c_{\mathrm{eff}}=0.4$.\vspace{1cm}} 
 \label{typical} 
 \end{floatingfigure}}
 A particularly well-known family of non-unitary CFTs are the non-unitary minimal models, commonly denoted by $\mathcal{M}_{p,q}$ with $c_{\mathrm{eff}}=1-\frac{6}{pq}$, $q>p+1$ and $p, q$ coprime.  RG flows between the non-unitary minimal models generated through perturbation by particular CFT fields have been investigated for a long time.  Early examples of such flows were pointed out in \cite{Lassig:1991an, Ahn:1992qi}. They proved that perturbation by a field $\phi_{1,3}$ (the least relevant field of the theory), generates the family of flows 
 \beq 
 \mathcal{M}_{p,q} + \phi_{1,3} \rightarrow \mathcal{M}_{2p-q,p}.
 \label{flow1}
 \eeq 
 The proof relied on perturbation theory under the assumption that $p/(q-p)\gg 1$. Further examples of RG flows amongst non-unitary minimal models have been found through the use of the thermodynamic Bethe ansatz (TBA) approach \cite{TBA1} and its massless version \cite{massless}. In particular, the families 
 \beqa 
&&  \mathcal{M}_{p,2p-1} + \phi_{2,1} \rightarrow \mathcal{M}_{p-1,2p-1}\nonumber \\
&&  \mathcal{M}_{p,2p+1} + \phi_{1,5} \rightarrow \mathcal{M}_{p,2p-1}
\label{flows2}
 \eeqa 
  were proposed in \cite{Martins:1991hi} and further studied in \cite{ravanini:1994pt}. As an example, the solid line in Fig.~\ref{typical} shows the massless flow $ \mathcal{M}_{3,5} + \phi_{2,1} \rightarrow \mathcal{M}_{2,5}$
  as presented in \cite{dorey:2000zb}.  A more general set of flows was proposed and explored in \cite{dorey:2000zb} giving the following families:
  \beqa 
&&  \mathcal{M}_{p,q} + \phi_{2,1} \rightarrow \mathcal{M}_{q-p,q} \quad \mathrm{for} \quad p<q<2p \nonumber \\
&&  \mathcal{M}_{p,q} + \phi_{1,5} \rightarrow \mathcal{M}_{p,4p-q} \quad \mathrm{for} \quad 2p<q<3p \nonumber \\
&&  \mathcal{M}_{p,q} + \phi_{1,5} \rightarrow \mathcal{M}_{4p-q,p}\quad \mathrm{for} \quad 3p<q<4p.
\label{flows3}
 \eeqa 
 which contain some of the examples above but are more general. 
 
 Within the TBA approach it is common to define a scaling function $c_{\mathrm{eff}}^{\mathrm{TBA}}(s)$ which for unitary theories has the same features as Zamolodchikov's $c$-function (even though they are distinct functions). The TBA scaling function has been evaluated for some of the flows listed above. In all cases the property (\ref{order}) is observed (and can be easily checked by simply evaluating $c_{\mathrm{eff}}$ for the listed theories). Moreover, in many examples, such as the flows (\ref{flows2}), the TBA scaling function is monotonic, although this is not always the case \cite{dorey:2000zb}. 
   
In addition to the massless flows (\ref{flow1})-(\ref{flows3}) between nontrival non-unitary CFTs, there are also known examples of massless flows connecting unitary to non-unitary minimal models. 
An example is provided by the QFT studied in \cite{fonseca} which may be described as implementing the flow 
\beq 
\mathcal{M}_{3,4}+ \lambda_ 1\phi_{1,3}+i  \lambda_2 \phi_{1,2} \mapsto \mathcal{M}_{2,5}
\eeq 
with $\lambda_{1,3}\in \mathbb{R}$. Here $\mathcal{M}_{3,4}$ is the critical Ising model, $\mathcal{M}_{3,4}+ \lambda_ 1\phi_{1,3}$ is the massive Ising model, $i\lambda_2\phi_{1,2}$ represents an imaginary magnetic field, and the final theory (the IR point) is the critical Lee-Yang model. Finally, there are also examples of massive flows from a non-unitary minimal model to the trivial fixed point,  $(c_{\mathrm{eff}})_{\mathrm{IR}}=0$. The dashed line in Fig.~\ref{typical} shows the massive flow 
\beq
\mathcal{M}_{2,5}+i \lambda \phi_{1,2} \mapsto \{{\bf 1}\}
\eeq 
with $\lambda\in\R$, from the minimal Lee-Yang model described by its massive integrable QFT counterpart \cite{Cardymuss,TCSA, Z}.

The examples above make it reasonable to expect that a generalization of Zamolodchikov's $c$-theorem will give rise to a similar-looking ``$c_{\mathrm{eff}}$-theorem''. In this letter we  show that a $c_{\mathrm{eff}}$-function exists with all the properties of Zamolodchikov's $c$-function.
This is shown under the standard QFT properties of Poincar\'e invariance and locality, as well as {\it unbroken parity--time-reversal ($\mathcal{PT}$) symmetry}. At critical points, under certain additional natural assumptions which hold at least in rational models of CFT, this function equals the effective central charge \cite{ceff}. This thus shows that the space of $\mathcal{PT}$-symmetric CFTs is partially ordered by ($\mathcal{PT}$-symmetric) RG flows, and that this order is characterized by \eqref{order}.

Interestingly, there also exist examples of flows where the inequality (\ref{order}) is violated \cite{nonmo1,nonmo2,nonmo25,nonmo3,nonmo4}. We discuss how these may break some of the assumptions underlying our result in Section \ref{sectest}.

The paper is organized as follows. In Section \ref{secgen} we recall some general facts about QFT in order to emphasize which results hold without unitarity, and we express the precise meaning of $\mathcal{PT}$-symmetry we use in order to replace unitarity. In Section \ref{sechash} we derive some basic consequences including a reflection positivity statement, and we discuss aspects of locality. In Section \ref{secthm} we prove our main irreversibility theorem, following closely the steps of Zamolodchikov's original proof. Finally in Section \ref{sectest} we briefly discuss various examples, and we conclude in Section \ref{secconclu}.


\section{Generalities and main assumptions}\label{secgen}

In this section, we establish the precise context in which we work, stating the assumptions and their immediate consequences. Except where otherwise stated, we work in real time with Minkowskian metric. There are three assumptions. The first two, Poincar\'e invariance and locality, are standard assumptions in relativistic QFT.  The last one is $\mathcal{PT}$-symmetry. If unbroken -- that is, if there is a basis of states that are $\mathcal{PT}$-invariant -- this symmetry guarantees the reality of the hamiltonian spectrum in non-hermitian quantum models \cite{wigner,BB}. In our proof of irreversibility, we need both reality of spectra and the more dynamical statement of $\mathcal{PT}$-invariance of the stress-energy tensor.

\subsection{Poincar\'e invariance and locality}

We consider a Hilbert space ${\cal H}$ on which we define a non-unitary quantum system. The Hilbert space has all appropriate structures, including a non-degenerate inner product which we denote with the usual Dirac bra-ket notation $\bra v|w\ket$, and an associated hermitian structure ${}^\dag$, with as usual $|v\ket^\dag = \bra v|$. The non-unitary quantum system is defined by its non-hermitian, diagonalizable hamiltonian $H$, with $H \neq H^\dagger$. Except for unitarity, other standard assumptions of relativistic QFT are made, which we review here for clarity.

We assume translation invariance, with associated momentum operator $P$ which satisfies
\beq
	[H,P]=0.
\eeq
Space and time translations are effected as usual as
\beq
	\Or(x,t) = e^{iHt-iPx}\Or e^{-iHt+iPx}
\eeq
where we identify $\Or = \Or(0,0)$.
As per the standard precepts of QFT, $H$ and $P$ are integrals of hamiltonian and momentum densities with locality properties:
\beq 
H=\int dx \, h(x,t), \quad P=\int dx \, p(x,t)
\eeq
with
\beq\label{hh}
[h(x,t),h(y,t)] = [h(x,t),p(y,t)] = [p(x,t),p(y,t)]=0\quad \forall\;x\neq y,\quad \forall t
\eeq
A (homogeneous) local field $\Or(x)=\Or(x,0)$ is an operator satisfying $[P,\Or(x)]=i\p_x\Or(x)$, and such that $[h(x),\Or(y)]=[p(x),\Or(y)]=0$ for all $x\neq y$. The usual locality arguments are assumed: let $\Or(x)$ be local; then if $\int \dd x\,\Or(x)=0$ then $\Or(x) = \p_x \t\Or(x)$ for local $\t\Or(x)$; and if $\p_x \Or(x) = 0$ then $\Or(x)= a {\bf 1}$ for some constant $a\in\C$. The above equations then imply the existence of local currents $j(x,t)$ and $k(x,t)$ such that
\beq\label{cur}
\partial_t {h}(x,t)+\partial_x {j}(x,t)=0, \quad 
\partial_t {p}(x,t)+\partial_x {k}(x,t)=0.
\eeq
As usual, these hold inside correlation functions except at the space-time positions where other local fields are inserted, in which case $\delta$-type contact terms arise, see e.g. \cite{DMS} or any QFT textbook.

We assume clustering of correlation functions, namely factorization of correlators of local fields at large space distances.

We further assume Lorentz invariance. The basic condition of Lorentz invariance is the equality between the energy current and the momentum density,
\beq
	j=p.
\eeq
Then we can construct the boost operator
\beq
	B = \int \dd x\,xh(x),
\eeq
which satisfies the correct relation for the (two-dimensional) Poincar\'e group,
\beq\label{bh}
	[B,H] = i\int \dd x\,x \p_th(x,t) = -i\int \dd x\,x\p_x p(x,t)
	= i\int \dd x\,p(x,t) = iP
\eeq
and
\beq\label{bp}
	[B,P] = -i\int \dd x\,x\p_x h(x,t) = i H.
\eeq
These relations hold up to local densities at infinity. By the clustering property, such local densities at infinity do not contribute whenever the operators are appropriately exponentiated, for instance when acting on local observables by adjoint action, hence can be neglected.

By using \eqref{bh}, \eqref{bp} as well as arguments based on locality, it is possible to deduce the following relations $i[B,h] = -2p$, $i[B,k]=-2p$ and $i[B,p]=-h-k$, up to derivatives of local fields. In the standard form of the stress-energy tensor, these derivatives are assumed to be zero (see a discussion of the stress-energy tensor in \cite{DMS}). Since such fundamental relations are not expected to be based on unitarity, here we assume that they hold.

Let $z=x-t$ and $\bar{z}=x+t$. By the Poincar\'e algebra, in general we have
\beq
	e^{i\alpha B}\Or(z,\b z)e^{-i\alpha B} = (e^{i\alpha B}\Or e^{-i\alpha B})
	(e^{-\alpha} z,e^{\alpha} \b z).
\eeq
Following the usual construction, consider the operators 
\beqa
{\tau}(z,\bar{z})&=&\frac{h(x,t)+k(x,t)+2p(x,t)}{4}\n \bar{\tau}(z,\bar{z})&=&\frac{h(x,t)+k(x,t)-2p(x,t)}{4}\n \theta(z,\bar{z})&=&k(x,t)-h(x,t).
\label{operators}
\eeqa
These have spin 2, -2 and 0 respectively,
\beq\label{btau}
	i[B,\tau] = -2\tau,\quad i[B,\b \tau] = 2\b\tau,\quad
	i[B,\theta]=0,
\eeq
from which we find the standard transformation properties,
\beq\label{Btrans}
	e^{i\alpha B}\tau e^{-i\alpha B} = e^{-2\alpha}\tau,\quad
	e^{i\alpha B}\b\tau e^{-i\alpha B} = e^{-2\alpha}\b\tau,\quad
	e^{i\alpha B}\theta e^{-i\alpha B} = \theta.
\eeq
These operators satisfy
\beq\label{curb}
	\partial \bar{\tau}+\frac{1}{4}\bar{\partial}\theta=0 \quad\mbox{and} \quad
	\bar{\partial}\tau +\frac{1}{4}\partial \theta=0
\eeq
where $\p = \p/\p z$ and $\b\p = \p/\p\b z$.

\subsection{$\mathcal{PT}$ symmetry and real spectra}

The core of our derivation of the $c_\mathrm{eff}$ theorem is the reality of the hamiltonian and momentum spectra, as well as $\mathcal{PT}$-invariance of the stress-energy tensor and of the ground state. For a $\mathcal{PT}$-invariant diagonalizable hamiltonian, the statement that all its eigenstates have real eigenvalues is equivalent to the statement that one can choose a basis of $\mathcal{PT}$-invariant eigenstates (that is, to the statement that $\mathcal{PT}$-symmetry is not spontaneously broken)\footnote{It is well-known that unbroken $\mathcal{PT}$-symmetry is not a necessary condition for the energy spectrum to be real. This can be guaranteed by the condition of pseudo-hermiticity as discussed at length in Ali Mostafazadeh's work \cite{Mo1,Mo2,Mo3}. However we would like to emphasize that our current derivation does require unbroken $\mathcal{PT}$-symmetry since, together with the reality of the spectrum, we require QFT correlators to be $\mathcal{PT}$-invariant (see section 4).}.
The simple idea that unbroken $\mathcal{PT}$-symmetry guarantees the reality of the energy spectrum even if a hamiltonian is non-hermitian is at the heart of an active area of research. The ideas underpinning this research go back to the early days of quantum mechanics \cite{wigner} when it was realized that hermiticity of the hamiltonian is not strictly required in order to define a meaningful quantum mechanical model. The area has been popularized more recently thanks to a great extent to the pioneering work \cite{BB} in which a family of non-hermitian quantum mechanical hamitonians  was shown to exhibit real energy spectrum. Various reviews of the field can be found in \cite{Carla,Bender,Mosta}. More recent developments are covered in the special issue \cite{spePT}.

Parity invariance is the transformation $x\rightarrow -x$, equivalently $z\leftrightarrow -\bar{z}$. Time reversal is the operation of complex conjugation of coefficients of states and operators as written in a chosen basis (the choice of the basis defines the time reversal operation). The combination is the operation $\mathcal{PT}$, which we will see (by a slight abuse of notation) as acting on operators and on vectors and co-vectors. We understand this operation as an anti-linear involution of the operator algebra, which preserves the inner product up to complex conjugation, $\mathcal{PT}(\bra v|)\mathcal{PT}(|w\ket) = \bra v|w\ket^*$, and which preserves the momentum operator,
\beq\label{PTP}
	\mathcal{PT}(P) = P.
\eeq

Our basic dynamical assumption is that {\em the stress-energy tensor is $\mathcal{PT}$-invariant}. That is, we assume
\beq\label{PTass}
	\mathcal{PT}(h(0,0)) = h(0,0),\quad
	\mathcal{PT}(p(0,0)) = p(0,0),\quad
	\mathcal{PT}(k(0,0)) = k(0,0).
\eeq
Thanks to \eqref{PTP} and anti-linearity this implies $\mathcal{PT}(h(x,0)) = h(-x,0)$, hence $\mathcal{PT}(H) = H$, and therefore
\beq\label{PTh}
	\mathcal{PT}(h(x,t)) = h(-x,-t),\quad
	\mathcal{PT}(p(x,t)) = p(-x,-t),\quad
	\mathcal{PT}(k(x,t)) = k(-x,-t).
\eeq

Consider the simultaneous right-eigenvalue equations for the hamiltonian and momentum operator, $H|R_n\ket = E_n|R_n\ket$ and $P|R_n\ket = p_n|R_n\ket$. We further assume that {\em all eigenvalues $E_n$ and $p_n$ are real}, or equivalently \cite{wigner,BB} that {\em all eigenstates $|R_n\ket$ are $\mathcal{PT}$-invariant} (unbroken $\mathcal{PT}$-symmetry).

We also assume, as usual in QFT, the set $\{E_n\}$ to be bounded from below, and the lowest-energy state $|R_0\ket$ to be unique; by appropriate identity-shift of $H$ and $P$, we choose it to have zero energy and momentum, $E_0=p_0=0$. Note that by Lorentz invariance in QFT, we must have $E_n \geq |p_n|$.

Denoting by $\bra L_n|$ the simultaneous left-eigenvectors for the hamiltonian and momentum operator,  we therefore have
\beq \label{Heq}
H \neq H^\dagger, \quad H |R_n\rangle= E_n |R_n\rangle \quad \mathrm{and} \quad \langle L_n|H= E_n \langle L_n|,\quad E_n\in\R
\eeq 
and
\beq\label{Peq}
	P|R_n\ket = p_n|R_n\ket  \quad \mathrm{and} \quad \langle L_n|P= p_n \langle L_n|,\quad p_n\in\R.
\eeq
It follows that $H^\dagger |L_n\rangle= E_n |L_n\rangle$ and $\langle R_n|H^\dagger= E_n \langle R_n|$. The vector $|R_n\ket$ and co-vector $\bra L_n|$ are the right and left eigenvectors, respectively, of the hamiltonian $H$. For $H$ non-hermitian we have in general that $|R_n \rangle^\dagger \neq \langle L_n|$, and the vectors $|R_n\rangle$ and $| R_m\ket$ are not orthogonal under the Hilbert space structure, $\langle R_m|R_n\rangle\neq \delta_{nm}$. However, we may construct our basis in such a way as to have
\beq
	\langle L_n| R_m\rangle  =\delta_{nm}.
\eeq
In the context of non-hermitian quantum mechanics this is termed a biorthogonal basis \cite{bi0, bi1,bi2}. In particular, we have
\beq \label{decomp}
{\bf{1}}=\sum_n |R_n \rangle \langle L_n|.
\eeq 
We choose the vectors $|R_m\ket$ and $|L_m\ket$ as such. Note that both are $\mathcal{PT}$-invariant.
 That is,
\beq
	\mathcal{PT}(|R_n\ket) = |R_n\ket,\quad
	\mathcal{PT}(\bra L_n|) = \bra L_n|.
\eeq

\section{The hash operation and reflection positivity}\label{sechash}

In unitary models, the hermitian conjugation of the Hilbert space guarantees a very important property, at the basis of the ordinary $c$-theorem: reflection positivity. For non-hermitian hamiltonians, this does not hold anymore. However, in  the theory of $\mathcal{PT}$-symmetric quantum mechanics, it is known that when $\mathcal{PT}$ symmetry is not broken, a similarity transformation exists which relates the hamiltonian to a hermitian counterpart\footnote{Such a similarity transformation also exists under the condition of pseudo-hermiticity \cite{Mo1,Mo2,Mo3}, distinct from $\mathcal{PT}$-symmetry and less stringent in some respects. However, as noted in footnote 2, we do require $\mathcal{PT}$-symmetry in the present context.} \cite{Carla,Bender,Mosta,spePT}.
With this new hermitian structure, the usual results of quantum mechanics hold. In the present context, it is important to specify how this new hermitian structure interacts with locality and Poincar\'e invariance.

\subsection{Hash operation}

Instead of exhibiting the explicit similarity transformation, we instead define the ${}^\hash$ operation on ${\rm End}({\cal H})$, which plays the role of the new hermitian conjugation. Let $\Or\in {\rm End}({\cal H})$ be an operator on the Hilbert space. Then $\Or^\hash$ is defined by the relation
\beq \label{hash}
\langle L_m| {\mathcal{O}}^{\hash} |R_n \rangle:= \langle L_n| {\mathcal{O}} |R_m \rangle^*=
\langle R_m| {\mathcal{O}}^\dagger |L_n \rangle
\eeq 
(where we recall that ${}^*$ is complex conjugation). It is a simple matter to see that the hash operation is an antilinear involution, and further that
\beq\label{o1o2}
	(\Or_1 \Or_2)^\hash = \Or_2^\hash \Or_1^\hash.
\eeq
The latter is derived as follows:
\beqa
	\langle L_m| (\mathcal{O}_1\mathcal{O}_2)^{\hash}
	|R_n \rangle
	&=&
	\langle L_n| \mathcal{O}_1 \mathcal{O}_2 |R_m \rangle^*
	=
	\lt(\sum_k \langle L_n| \mathcal{O}_1|R_k\ket \bra L_k|
	\mathcal{O}_2 |R_m \rangle\rt)^* \n
	&=&
	\sum_k \langle L_k| \mathcal{O}_1^\hash|R_n\ket \bra L_m|
	\mathcal{O}_2^\hash |R_k \rangle
=
	 \langle L_m| \mathcal{O}_2^\hash
	\mathcal{O}_1^\hash |R_n \rangle. 
\eeqa
Also, thanks to reality of spectra, we find that $\langle L_m| {H}^{\hash} |R_n \rangle= \langle R_m| {H}^\dagger |L_n \rangle=E_n \delta_{m,n}=\langle L_m| {H} |R_n \rangle$ and similarly for $P$, and thus
\beq\label{hphash}
	{H}^{\hash}=H\quad\mathrm{and}\quad P^\hash = P.
\eeq
In particular,
\beq\label{oxt}
	\Or(x,t)^\hash = (e^{iHt-iPx}\Or(0,0)e^{-iHt+iPx})^\hash
	= e^{iHt-iPx}\Or(0,0)^\hash e^{-iHt+iPx}
	= \Or^\hash(x,t).
\eeq

The hash operation may also be defined as an antilinear operation from $\mathcal H$ to its dual $\mathcal H^*$, and vice versa. Consistency with the above definition on operators implies
\beq
	|R_n\ket^\hash = \bra L_n|,\quad
	\bra L_n|^\hash = |R_n\ket.
\eeq
From this and the fact that the basis states $|R_n\ket$ and $\bra L_n|$ are $\mathcal{PT}$-invariant, we see that $^\hash$, as acting on $\mathcal H$ and as acting $\mathcal H^*$, commutes with $\mathcal{PT}$. Therefore, it does so as well as acting on ${\rm End}(\mathcal{H})$:
\beq\label{PThash}
	\mathcal{PT}(\Or^\hash) = \mathcal{PT}(\Or)^\hash.
\eeq

\subsection{Reflection  positivity}

Let us now consider a generic ground-state two-point function $\langle L_0| {\Or}_1(x_1,t_1) \Or_2(x_2,t_2)|R_0\rangle$ of some operators $\Or_1,\, \Or_2$. 
Using the decomposition of the identity \eqref{decomp} as well as \eqref{Peq} and \eqref{Heq}, we find that it is a function of time and position differences only:
\beq 
\langle L_0| {\Or}_1(x_1,t_1) \Or_2(x_2,t_2)|R_0\rangle=\sum_n e^{-i(t_1-t_2)E_n + i(x_1-x_2)p_n} \langle L_0| {\Or}_1| R_n\rangle \langle L_n| \Or_2|R_0\rangle.
\eeq
Using \eqref{hash}, \eqref{o1o2} and \eqref{oxt}, we also see that
\beq 
\langle L_0| {\Or}_1(x_1,t_1) \Or_2(x_2,t_2)|R_0\rangle=\langle L_0| {\Or}_2^{\hash}(x_2,t_2) \Or_1^{\hash}(x_1,t_1)|R_0\rangle^* .
\label{rel1}
\eeq

Finally, consider the time-dependent two-point function of ${\mathcal{O}}^{\hash}$ and $\mathcal{O}$ in imaginary time $t=-iy,\;y\in\R$, in the ground state of the theory,
\beq 
\langle L_0|{\mathcal{O}}^{\hash} (x,-iy) \mathcal{O}(x,0) |R_0\rangle=\langle L_0|e^{-y H}{\mathcal{O}}^{\hash} (x,0)e^{y H} \mathcal{O}(x,0) |R_0\rangle=\langle L_0|{\mathcal{O}}^{\hash}(x,0)e^{y H} \mathcal{O}(x,0) |R_0\rangle.
\eeq 
Using \eqref{decomp} we write
\beq 
\langle L_0|{\mathcal{O}}^{\hash}(x,-iy) \mathcal{O}(x,0) |R_0\rangle=\sum_n e^{y E_n} \langle L_0| {\mathcal{O}}^{\hash}(x,0) |R_n \rangle \langle L_n| {\mathcal{O}}(x,0) |R_0 \rangle. 
\eeq 
Thanks to \eqref{hash}, this sum is {\it positive-(semi)definite}:
\beq\label{pos}
	\langle L_0|{\mathcal{O}}^{\hash}(x,-iy) \mathcal{O}(x,0) |R_0\rangle \geq0
\eeq
(equality occurs for some value of $y$, if and only if all correlation functions involving $\Or(x,t)$ vanish). This is reflection positivity.

\subsection{Hash-locality and conserved currents}

Since $H$ and $P$ are invariant under the hash operation, it is clear that the relations defining locality (see the discussion below \eqref{hh}) may be hashed, keeping invariant the position $x$. We may therefore distinguish two classes of locality, both associated to the same real space parameterized by $x$: {\em locality} and {\em hash-locality}, the first with respect to $h(x)$ and $p(x)$, the second to $h^\hash(x)$ and $p^\hash(x)$. Hashing \eqref{cur}, we have
\beq\label{curhash}
\partial_t {h}^{\hash}(x,t)+\partial_x {j}^{\hash}(x,t)=0, \quad 
\partial_t {p}^{\hash}(x,t)+\partial_x {k}^{\hash}(x,t)=0,
\eeq 
where all fields are hash-local. Relations \eqref{curhash} hold in correlation functions except at space-time positions $(x,t)$ at which other hash-local fields are inserted, where additional standard contributions in the form of $\delta$-function contact terms will arise. Passing to the $\tau^\hash$, $\b\tau^\hash$ and $\theta^\hash$ fields, these equations are
\beq\label{hashcurb}
	\partial \bar{\tau}^\hash+\frac{1}{4}\bar{\partial} \theta^\hash=0\quad\mbox{and} \quad
	\bar{\partial}\tau^\hash +\frac{1}{4}\partial \theta^\hash=0.
\eeq

It will be important below to consider mixed correlation functions, $\bra L_0|\Or_1^\hash(x,t)\Or_2(0,0)|R_0\ket$ for local $\Or_2$ and hash-local $\Or_1^\hash$. Clearly, local fields are not necessarily hash-local, and can be hash-supported on extended regions. Yet, since conservations equations hold if the fields are time-separated, relations \eqref{curhash} still hold in correlation functions with insertion of local fields at times different from $t$.


\subsection{Spin}

By hashing \eqref{btau}, it is clear that the hash-local fields $\tau^\hash$, $\b\tau^\hash$ and $\theta^\hash$ have hash-spin 2, $-2$ and 0. However, in order to determine the space-time dependence of two-point functions involving a mixture of local and hash-local fields, we need to have a common notion of spin for both types of locality. Thanks to $H^\hash=H$ and $P^\hash=P$, the Poincar\'e algebra implies $[H,B-B^\hash]=[P,B-B^\hash]=0$. It is in fact natural to assume hash-invariance of all Poincar\'e generators, imposing $B^\hash=B$ and thus a single notion of spin. In order to justify this assumption, we provide general arguments that indeed lead to $B^\hash=B$ for the boost operator.

Let us parametrize the energy and momentum eigenvalues of a right eigenstate $|R_n\rangle$ by $m_n\cosh\theta_n$ and $m_n\sinh\theta_n$ respectively\footnote{Note that although this parametrization may seem reminiscent of two-dimensional QFT where $m_n$ typically represents the mass and $\theta_n$ the rapidity of a one particle excitation, here $m_n$ and $\theta_n$ are introduced as generic parameters where $m_n$ is just the minimum energy (corresponding to the zero-momentum state) and $\sinh \theta_n$, $\cosh\theta_n$ simply arise from a Lorentz boost of the associated minimal energy state.}, for some $m_n > 0$ and $\theta_n\in\R$. This can always be done by positivity of the $H$ spectrum and reality of the $P$ spectrum if $E_n>|p_n|$; in the case $E_n=|p_n|$ (``massless states"), an alternative parametrization can be used, without affecting the argument\footnote{Massless states can be parameterized using the exponential function $me^\theta$ instead, for an arbitrarily chosen $m$. For such states, a boost cannot bring the momentum to zero, and cannot change its sign. Instead, the two values $\pm m$ of momentum can be used as anchors, one positive and one negative.}. Its boost $e^{i\alpha B}|R_n\ket$ by the rapidity $\alpha\in\R$ also is a right eigenstate, with energy and momentum eigenvalues $m_n\cosh(\theta_n+\alpha)$ and $m_n\sinh(\theta_n+\alpha)$ respectively. Therefore, we may always bring its momentum to 0.  The same holds for the boost of the left eigenstate $\bra L_n|e^{-i\alpha B^\hash}$.

Let us change the labeling, and use the symbols $|R_n,0\ket$ in order to label a basis of independent center-of-momentum right eigenstates and $\bra L_n,0|$ the corresponding orthogonal left eigenstates, $\bra L_n,0|R_m,0\ket = \delta_{m,n}$ and $P|R_n,0\ket = 0$, $\bra L_n,0|P = 0$.
Let us define the states $|R_n,\alpha\ket := e^{i\alpha B}|R_n,0\ket$ and $\bra L_n,\alpha|:=\bra L_n,0|e^{-i\alpha B^\hash}$ for all $n$ and all $\alpha\in\R$. By the above discussion the set of states $\{|R_n,\alpha\ket\}$ span the Hilbert space, and $\{\bra L_n,\alpha|\}$ spans its dual. Let us now consider the Fourier transforms
\beq
	|R_n,\omega\ket\ket = \int d\alpha\,e^{-i\omega\alpha}|R_n,\alpha\ket,\quad
	\bra\bra L_n,\omega| = \int d\alpha\,e^{i\omega\alpha}\bra L_n,\alpha|
\eeq
for $\omega\in\R$. Note that, since real boosts $\alpha\in\R$ are used, it is important that $\omega$ be real for existence of the integrals\footnote{Overlaps $\bra L_n,\alpha|R_m,\beta\ket$ may be chosen of the form $f(\alpha)\delta(\alpha-\beta)\delta_{m,n}$. The argument presented here requires $f(\alpha)$ to grow at most polynomially at large $|\alpha|$, and the conclusion is that $f(\alpha)$ is in fact independent of $\alpha$.}. Since the Fourier transform is invertible, these states span the Hilbert space and its dual respectively, or at least dense subsets of these. Since $B|R_n,\alpha\ket = -i\p_\alpha|R_n,\alpha\ket$ and $\bra L_n,\alpha|B^\hash = i\p_\alpha\bra L_n,\alpha|$, they right- and left-diagonalize $B$ and $B^\hash$ respectively,
\beq
	B|R_n,\omega\ket\ket = \omega |R_n,\omega\ket\ket,\quad
	\bra\bra L_n,\omega| B^\hash = \omega \bra\bra L_n,\omega|.
\eeq
Therefore,
\beq
	\bra\bra L_n,\omega|B|R_m,\nu\ket\ket = \omega
	\bra\bra L_n,\omega|R_m,\nu\ket\ket =
	\bra\bra L_n,\omega|B^\hash|R_m,\nu\ket\ket
\eeq
for all $n,m$ and $\omega,\nu$. Since the vectors and covectors span (dense subsets of) the Hilbert space and its dual, we conclude that $B=B^\hash$.

As a consequence,
\beq\label{Btranshash}
	e^{i\alpha B}\tau^\hash e^{-i\alpha B} = e^{-2\alpha}\tau^\hash,\quad
	e^{i\alpha B}\b\tau^\hash e^{-i\alpha B} = e^{-2\alpha}\b\tau^\hash,\quad
	e^{i\alpha B}\theta^\hash e^{-i\alpha B} = \theta^\hash.
\eeq

\section{Irreversibility theorem}\label{secthm}

The irreversibility theorem is composed of two parts. The first part is the proof that a certain function $c_\mathrm{eff}(s)$ monotonically decreases along the RG flow (where $s$ is the log of the RG distance scale). This follows very closely Zamolodchikov's proof \cite{zamc}. The function is defined in terms of two-point functions involving stress-energy tensor components and their hash counterparts. Reflection positivity, $\mathcal{PT}$-symmetry, and the fact that the Poincar\'e group is hash-invariant ($H^\hash=H$, $P^\hash=P$ and $B^\hash=B$) are used in a fundamental way. The function strictly decreases from its initial (UV) to its final (IR) point, unless the trace of the stress-energy tensor $\theta(x,t)$ vanishes, in which case the function is constant.

The second part aims at identifying $c_\mathrm{eff}(s)$ at fixed points. This is based on an independent argument, using the critical specific free energy. It is known that, in many models of CFT including all rational models, the specific free energy is proportional to the CFT effective central charge $c_\mathrm{eff} = c-24\Delta$, where $c$ is the central charge and $\Delta$ is the lowest dimension of the spectrum of local fields (in non-rational, non-compact models the choice of $\Delta$ may be more delicate). Thus, we identify
\beq
	c(\infty) = c_\mathrm{IR}-24\Delta_\mathrm{IR}\qquad\mbox{and}\qquad c(-\infty) = c_\mathrm{UV}-24\Delta_\mathrm{UV}.
\eeq

\subsection{A monotonic function}

We first define the operation of taking the ``real part'', or hash-invariant part, of an operator:
\beq
	A^\real = \frc{A+A^\hash}2.
\eeq
Clearly $(A^\real)^\hash = A^\real$.

In the spirit of Zamolodchikov's original proof of the $c$-theorem,  we then define the correlators:
\begin{eqnarray}
&&f(z\bar{z}):= z^4 \langle L_0|  \tau^\real(z, \bar{z}) \tau^\real(0,0)|R_0 \rangle,\qquad \,\, \,\,\,\bar{f}(z\bar{z}):= \bar{z}^4 \langle L_0|  \bar{\tau}^\real(z, \bar{z}) \bar{\tau}^\real(0,0)|R_0 \rangle,\nonumber\\
&&g_1(z\bar{z}):=z^3 \bar{z}\langle L_0| \tau^\real(z, \bar{z})\theta^\real(0,0) |R_0\rangle,  \qquad \bar{g}_1(z\bar{z}):=\bar{z}^3 {z}\langle L_0| \bar{\tau}^\real(z, \bar{z})\theta^\real(0,0) |R_0\rangle, \nonumber\\
&& g_2(z\bar{z}):=z^3 \bar{z}\langle L_0| \theta^\real(z, \bar{z})\tau^\real(0,0) |R_0\rangle,  \qquad 
\bar{g}_2(z\bar{z}):=\bar{z}^3 {z}\langle L_0| {\theta}^\real(z, \bar{z})\bar{\tau}^\real(0,0) |R_0\rangle, \nonumber\\
&& q(z\bar{z}):= z^2 \bar{z}^2 \langle L_0|  \theta^\real (z, \bar{z})\theta^\real(0,0) |R_0 \rangle.
\label{fgh2}
\end{eqnarray}
The fact that all these functions depend on $z\b z$ is a consequence of \eqref{Btrans} and \eqref{Btranshash}. Here and below we consider space-like distances $z\b z>0$. Thanks to \eqref{curb} and \eqref{hashcurb}, there are various relations between the derivatives of these correlators, which hold for all $z\neq \b z$:
\beq 
\bar{z}\bar{\partial} f+\frac{z}{4} \partial g_2=\frac{3}{4} g_2 \quad \mathrm{and} \quad {z}{\partial} \bar{f}+\frac{\bar{z}}{4} \bar\partial \bar{g}_2=\frac{3}{4} \bar{g}_2,
\eeq 
as well as
\beq 
\bar{z}\bar{\partial} g_1+\frac{z}{4} \partial q=g_1+\frac{1}{2}q \quad \mathrm{and} \quad {z}{\partial} \bar{g}_1+\frac{\bar{z}}{4} \bar{\partial} q=\bar{g}_1+\frac{1}{2}q.
\eeq 
These are identical to the relations found by Zamolodchikov if we identify $g_{1}=\bar{g}_1=g_2=\bar{g}_2$ and $f=\bar{f}$. However, in general these identifications do not all hold in the present case.  
We may combine the relations above to write:
\beq 
\bar{z}\bar{\partial} f+{z}{\partial} \bar{f}+\frac{z}{4} (\partial g_2-3\partial \bar{g}_1)+\frac{\bar{z}}{4} (\bar\partial \bar{g}_2-3\bar\partial {g}_1)-\frac{3}{16} (z\partial+\bar{z}\bar{\partial}) q=\frac{3}{4}(g_2-g_1+\bar{g}_2-\bar{g}_1-q).
\eeq 
We may now use $\mathcal{PT}$ symmetry to show that $g_1=g_2$ and $\bar{g}_1=\bar{g}_2$. By translation invariance, hash-invariance of $\tau^\real$ and $\theta^\real$, and \eqref{rel1}, it follows that
\beq
\langle L_0| \tau^\real(z,\b z)\theta^\real(0,0) |R_0\rangle
=\langle L_0| \tau^\real(0,0)\theta^\real(-z, -\bar{z}) |R_0\rangle=\langle L_0| \theta^\real(-z,-\bar{z})\tau^\real(0, 0) |R_0\rangle^*.
\eeq 
Using $\mathcal{PT}$-symmetry \eqref{PTh} and \eqref{PThash}, we have
\beqa
 \langle L_0| \theta^\real(-z,-\bar{z})\tau^\real(0, 0) |R_0\rangle^*
&=&
\mathcal{PT}(\bra L_0|)\,\mathcal{PT}(\theta^\real(-z,-\b{z}))\,\mathcal{PT}(\tau^\real(0, 0))\,\mathcal{PT}(|R_0\ket) \n
&=& 
\bra L_0|\theta^\real(z,\b{z})\tau^\real(0, 0)|R_0\ket.
\eeqa
This implies $g_1=g_2$.  The same reasoning may be applied to $\bar{g}_1$ to show that $\bar{g}_1=\bar{g}_2$. Using this, the equation above simplifies to:
\beq 
\bar{z}\bar{\partial} f+{z}{\partial} \bar{f}+\frac{z}{4} (\partial g_1-3\partial \bar{g}_1)+\frac{\bar{z}}{4} (\bar\partial \bar{g}_1-3\bar\partial {g}_1)-\frac{3}{16} (z\partial+\bar{z}\bar{\partial}) q=-\frac{3}{4}q.
\eeq 
We may now change variables to polar coordinates by writing $z=r e^{\theta}$, $\bar{z}=r e^{-\theta}$, with $r>0$. We then have that $2 z \partial = r\partial_r + \partial_\theta $ and $2\bar{z}\bar\partial=r\partial_r- \partial_\theta$, and:
\beq 
\frac{1}{2} r\partial_r\left(f+\bar{f}-\frac{1}{2}(g_1+\bar{g}_1)-\frac{3}{8} q\right)- \frac{1}{2} \partial_\theta \left(\bar{f}-f +\bar{g}_1-g_1\right)=-\frac{3}{4}q.
\eeq 
Given that all functions involved are functions of $z\bar{z}=r^2$ only, it follows that the $\theta$-derivative must be zero, thus the equation simplifies to:
\beq 
 r\partial_r\left(f+\bar{f}-\frac{1}{2}(g_1+\bar{g}_1)-\frac{3}{8} q\right)=-\frac{3}{2}q,
\eeq 
or, introducing the standard RG parameter $s=2\log(mr)$ (where $m$ is an energy scale) we can write:
\beq 
 \frac{d}{ds}\left(f+\bar{f}-\frac{1}{2}(g_1+\bar{g}_1)-\frac{3}{8} q\right)=-\frac{3}{4}q.
\eeq 
This now takes almost exactly the same form as Zamolodchikov's $c$-theorem.

We may now define a function
\beq 
c_{\mathrm{eff}}(s):= 4\pi^2\left( f(r)+\bar{f}(r)-\frac{1}{2}(g_1(r)+\bar{g}_1(r))-\frac{3}{8}q(r)\right),
\label{defcef}
\eeq
which satisfies the equation
\beq\label{dc}
\frac{dc_{\mathrm{eff}}}{ds}=-3\pi^2 q(r).
 \eeq 
 The factor $(2\pi)^2$ in (\ref{defcef}) is introduced in order to reestablish the standard conformal normalization of the fields $\tau, \bar{\tau}$ and $\theta$ which, as shown in the paragraph before equation (\ref{7}), usually involves an extra factor $-2\pi$. 
 
Consider the function $q(r)$. Its values for all $r>0$ may be obtained from the analytic continuation of the correlator $\bra L_0|\theta^\real(z,\b z)\theta^\real(0,0)|R_0\ket$ to purely imaginary times $t=-iy$ (recall that $z=x-t$ and $\b z=x+t$). The inequality $q(r) \geq 0$ follows from reflection positivity \eqref{pos}. Thus, if $c_{\mathrm{eff}}(s)$ is complex, then its imaginary part is in fact independent of $s$, and its real part is strictly monotonically decreasing except when $q(r)$ is zero.  By \eqref{pos}, it is clear that $q(r)$ vanishes for some $r$ if and only if it does so for all $r>0$, and this happens if and only if  $\bra L_n|\theta|R_0\ket=0$ for all $n$. In this case all vacuum correlation functions involving the trace of the stress-energy tensor vanish, and thus we may set it to zero. That is, either the real part of $c_\mathrm{eff}(s)$ is strictly monotonically decreasing as $s$ increases from $-\infty$ to $\infty$, or it is constant for all $s$ and $\theta(x,t)=0$.

As usual, we assume that the limits $\lim_{s\to\pm\infty}c_\mathrm{eff}(s)$ exist, and that these correspond to the UV and IR quantum critical points, at which scale invariance holds and thus the trace of the energy-momentum tensor vanishes. At these points, $g_1=\b g_1=0$. By analytic continuation to imaginary times, one shows by similar arguments as above that $f$ and $\b f$ are real and non-negative. Therefore, $c_\mathrm{eff}(\pm\infty)\geq 0$, and combining with \eqref{dc}, this implies that $c_\mathrm{eff}(s)\geq 0$ 

Thus we have established the existence of a function $c_{\mathrm{eff}}(s)$ of the renormalization group parameter $s$ which is non-negative and monotonically decreasing along renormalization group flows, with
\beq
	c_\mathrm{eff}(-\infty) \geq  c_\mathrm{eff}(\infty)
\eeq
(there is equality if and only if $\theta(x,t)=0$, in which case $c_\mathrm{eff}(s)$ is independent of $s$). This is irreversibility of the RG flow.

The only missing bit of the puzzle is the determination of the values $c_{\mathrm{eff}}(\pm\infty)$, at the IR and UV quantum critical points. From the definitions above these are given by the correlator $4\pi^2(f(r_*)+\bar{f}(r_*))$ at $r_*=0$ or $r_*=\infty$. In contrast to the unitary case, however, we do not immediately know these values as we do not have explicit expressions for the operators $\tau, \tau^\hash$ within the standard CFT framework. We thus need to resort to a different strategy.

\medskip
\noindent {\em Remark.} In fact, even for unitary theories, the result above is slightly more general, as it does not assume that  $f=\bar{f}$ and $g_1=\bar{g}_1$. Indeed, there exist CFTs (even unitary ones) where the left and right central charges may be different. In such cases $f\neq \bar{f}$ and $g_1\neq \bar{g}_1$. As expected even for those theories there is an irreversibility theorem, which, following the reasoning here, would be a ``$c+\bar{c}$-theorem" where the value of the scaling function at critical points would be $\frac{c+\bar{c}}{2}$.  Having $c\neq \bar{c}$ means that the theory is not separately parity and time-reversal invariant, even if it is $\mathcal{PT}$-symmetric\footnote{Note that modular invariance implies that,  even in cases where $c\neq \bar{c}$, the difference $c-\bar{c}$ is constrained to take values in $12\mathbb{Z}$ \cite{DMS}.}. 

\medskip
\noindent {\em Remark.} Different choices of the functions $f, \b f, g_i, \b g_i$ and $q$ are possible. For instance, we could have used $f(z\b z) = z^4 \bra L_0|\tau^\hash(z,\b z)\tau(0,0)|R_0\ket$, etc., with the same result. These may simply be different monotonic functions along the RG flow. For the argument presented in the next section, the choice used here is more convenient.

\subsection{Connection with the CFT effective central charge}

In order to evaluate the values of $c_\mathrm{eff}(\pm\infty)$, we need to calculate the function $4\pi^2(f(r)+\bar{f}(r))$ in a CFT. The statement of scale invariance is the vanishing of the trace of the stress-energy tensor,
\beq
	\theta(x,t)=0.
\eeq
From this alone, it is possible to conclude that chiral factorization occurs, and that $f(r)$ and $\b f(r)$ are constant. In order to emphasize that these arguments do not depend on unitarity, we repeat them briefly here.

First, equations \eqref{curb} and the similar relations for hash-fields imply that in any correlation function, $\tau$ and $\tau^\hash$ are solely functions of $z$, and $\b\tau$  and $\b\tau^\hash$ functions of $\b z$ (this is true, as usual, except at the space-time positions of other local or hash-local field insertions). Second, suppose a state $\bra\cdots\ket$ is space-time translation invariant and clustering at large distance. Then,
\beqa
	&& \bra \tau(x,0) \b\tau(x',0)\ket
	= \lim_{t\to\infty} \bra\tau(x,t) \b\tau(x',t)\ket \n
	&& = \lim_{t\to\infty} \bra  \tau(x-t,0) \b\tau(x'+t,0)\ket
	= \bra \tau(0,0)\ket\,\bra \b\tau(0,0)\ket.
	\label{chiral}
\eeqa
This is (a part of) chiral factorization. Third, using the fact that $\bra L_0|\tau^\real(z,\b z)\tau^\real(0)|R_0\ket$ (resp. $\bra L_0|\b \tau^\real(z,\b z)\b \tau^\real(0)|R_0\ket$) only depends on $z$ (resp. $\b z$) for all $z\neq\b z$, and that the ground state is Lorentz invariant, the unique one-parameter solutions to \eqref{Btrans} are
\beq\label{gsres}
	\bra L_0|\tau^\real(z,\b z)\tau^\real(0)|R_0\ket = A z^{-4},
	\quad
	\bra L_0|\b \tau^\real(z,\b z)\tau^\real(0)|R_0\ket = \b A\b z^{-4}
\eeq
for some constants $A$, $\b A$. This shows that $f(r)$ and $\b f(r)$ are indeed constants.

In the following, we will use the standard notation $\tau(z) = \tau(z,\b z)$ and $\b\tau(\b z) = \b \tau(z,\b z)$ (similarly for hashed fields) in order to emphasize chirality. We will also understand the variables $z$ and $\b z$ as complex variables (complex conjugate of each other), and use Euclidean, imaginary-time fields. Therefore $\tau(z)$ is holomorphic and $\b \tau(z)$ is anti-holomorphic, except at positions of other fields insertions, where singularities may occur.

Consider the partition function of a CFT at finite temperature $T=\beta^{-1}$ in a system of length $\ell$,
\beq 
Z=\sum_n e^{-\beta E_n(\ell)}.
\eeq 
It was shown in \cite{BCN,Affleck}, and then generalized to non-unitary models \cite{ceff} that, at least for minimal models of CFT, the specific free energy is given by:
\beq \label{Z}
\lim_{\ell \rightarrow \infty} \ell^{-1} \log Z=f_0\beta +\frac{\pi(c_{\mathrm{eff}}+\bar{c}_{\mathrm{eff}})}{12 \beta}
\eeq 
where $f_0$ is an energy per unit length, and
\beq\label{ceffcft}
	c_\mathrm{eff} = c-24\Delta,\quad \b c_\mathrm{eff} = \b c-24\b\Delta,
\eeq
with $c$ (resp. $\b c$) the holomorphic (resp. antiholomorphic) central charge and $\Delta$ (resp. $\b\Delta$) the lowest holomorphic (resp. antiholomorphic) dimension of the CFT.
In fact, in \cite{ceff} it was assumed that $c_{\mathrm{eff}}=\bar{c}_{\mathrm{eff}}$ as this holds for most CFTs, but the above is a simple generalization. More generally, relation \eqref{Z} is expected to hold simply based on scale invariance, and defines the quantity $c_{\mathrm{eff}}+\bar{c}_{\mathrm{eff}}$, which in non-compact models might or might not be determined by \eqref{ceffcft}.

For a non-unitary CFT we can write 
\beq 
Z=\sum_n \langle L_n | e^{-\beta H}| R_n\rangle. 
\eeq 
Differentiating with respect to $\beta$ twice we find
\beq 
\frac{\partial^2 Z}{\partial \beta^2}=\langle H^2 \rangle_\beta^\mathrm{c}
\eeq 
where $\bra AB\ket_\beta^\mathrm{c} = \bra AB\ket_\beta- \bra A\ket_\beta\bra B\ket_\beta$ and
\beq 
\langle \mathcal{O} \rangle_\beta= Z^{-1}{\sum_n\langle L_n|\mathcal{O}e^{-\beta H}|R_n\rangle}.
\eeq 
At criticality, we have $h = \tau + \b \tau$ and $p=\tau-\b\tau$, and thanks to chiral factorization \eqref{chiral},
 \beqa 
&&  \langle (\tau(z)+ \bar{\tau}(\bar{z}))(\tau(z')+\bar{\tau}(\bar{z}')) \rangle_\beta-\langle \tau(z)+ \bar{\tau}(\bar{z})\rangle_\beta\, \langle \tau(z')+\bar{\tau}(\bar{z}')\rangle_\beta\nonumber\\
&&= \langle \tau(z) \tau(z')\rangle_\beta - \langle \tau\rangle^2_\beta + \langle \bar{\tau}(\bar{z}) \bar{\tau}(\bar{z}')\rangle_\beta - \bra \b\tau\ket_\beta^2.
 \eeqa 
Therefore, defining $H_\pm = (H\pm P)/2$,
\beq
	\bra H^2\ket_\beta^\mathrm{c}
	= \bra H_+^2\ket_\beta^\mathrm{c}
	+ \bra H_-^2\ket_\beta^\mathrm{c}.
\eeq
Using the fact that ${H}^{\hash}=H$ and $P^\hash=P$ we may also write
\beq 
\frac{\partial^2 Z}{\partial \beta^2}=
	\bra H_+^\real H_+^\real  \ket_\beta^\mathrm{c}+
	\bra H_-^\real  H_-^\real \ket_\beta^\mathrm{c}
\eeq
and so
\beqa 
\frac{\partial^2}{\partial \beta^2} \left(\lim_{\ell \rightarrow \infty} \ell^{-1}\log Z\right)&=&\int dx \, \left(\langle {\tau}^\real(x,t+i\epsilon) \tau^\real (0,t)\rangle_\beta^\mathrm{c}+\langle \b\tau^\real (x,t+i\epsilon) \b\tau^\real (0,t)\rangle_\beta^\mathrm{c}\right)\nonumber\\
&=&\frac{\pi(c_{\mathrm{eff}}+\bar{c}_{\mathrm{eff}})}{6\beta^3}
 \eeqa 
where we shift the time variable slightly to ensure that operators are time-ordered.

In unitary CFT we may compute the correlators involved by identifying the CFT at finite temperature with a CFT on a cylinder of radius $\beta^{-1}$, and then employing the transformation properties of the energy-momentum tensor under a conformal map from the plane to the cylinder. However, once more, we have not identified correlation functions of our fields $\tau, \tau^\hash, \bar{\tau}, \bar{\tau}^\hash$ with those of the Virasoro-generated holomorphic and anti-holomorphic energy-momentum tensor of the standard formulation of CFT, hence we have not shown their transformation properites under conformal maps. We may instead use a more general QFT result, namely the Kubo-Martin-Schwinger relation \cite{kubo,MS}
 \beq \label{kms}
\langle {\tau}^{\real}(z) \tau^\real ({z}')\rangle_\beta=\langle {\tau}^\real (z') \tau^{\real}(z-i\beta)\rangle_\beta.
\eeq

In order to go further, we need to argue that
\beq\label{comm}
	\langle {\tau}^\real (z') \tau^{\real}(z)\rangle_\beta
	=\langle \tau^{\real}(z){\tau}^\real (z')\rangle_\beta.
\eeq
Consider the function $F(z,\beta) = \bra \tau^\real(z)\tau^\real(0)\ket_\beta$. It is analytic in a neighborhood of the line ${\rm Im}(z)=0$, except possibly at ${\rm Im}(z)=0$. On the one hand, for $z\approx 0$, by scaling one can use the ground-state result \eqref{gsres}, and thus there must be an isolated pole of order 4 at the origin. On the other hand, one can argue that the position of any other singularity, which might appear for instance due to the non-locality of $\tau^\hash$, cannot depend on the temperature, as it is a property of the operators, not of the state. Therefore, by scaling again, no other singularity should exist on the real line. By imaginary-time ordering, for $z\in\R\setminus\{0\}$ we have $F(z+i0) = \bra \tau^\real(z)\tau^\real(0)\ket_\beta$ and $F(z-i0) = \bra \tau^\real(0)\tau^\real(z)\ket_\beta$. Since $F(z)$ is analytic, hence continuous, on $\R\setminus\{0\}$, this implies \eqref{comm} on $z-z'\in\R\setminus\{0\}$, and therefore for all $z,z'$ by analytic continuation. Note that relation \eqref{comm} points to the equivalence, at least from the viewpoint of the stress-energy tensor and at criticality, of locality and hash-locality.

Combining \eqref{kms} and \eqref{comm} we obtain the statement of periodicity for the analytic function $F(z,\beta) = \bra \tau^\real(z)\tau^\real(0)\ket_\beta$:
\beq\label{per}
	F(z,\beta) = F(z+i\beta,\beta).
\eeq
The function is in fact expected to be analytic within the full strip ${\rm Im}(z)\in(-\beta/2,\beta/2]$ except for the pole at $z=0$. By general results in one-dimensional models at nonzero temperature, the two point function vanishes exponentially at large distances. The unique family of solutions to the periodicity \eqref{per}, the requirement of a singularity \eqref{gsres} at $z=0$, and exponential vanishing at larges distances is
 \beq 
\langle {\tau}^{\real}(z) \tau^{\real}({z}')\rangle_\beta=\frac{\pi^4 A}{\beta^4 \sinh^4\frac{\pi (z-z')}{\beta}},
 \eeq 
 where $A$ is a constant to be determined. Similarly:
  \beq 
\langle \bar{\tau}^{\real}(\bar{z}) \bar{\tau}^{\real}(\bar{z}')\rangle_\beta=\frac{\pi^4 \bar{A}}{\beta^4 \sinh^4\frac{\pi (\bar{z}-\bar{z'})}{\beta}}.
 \eeq 
 Performing the integral
\beq 
\frac{\pi^4}{\beta^4}\int dx \, \left(\frac{A}{\sinh^4\frac{\pi (x+i \epsilon)}{\beta}}+\frac{\bar{A}}{\sinh^4\frac{\pi (x-i\epsilon)}{\beta}}\right)=\frac{4\pi^3}{3\beta^3}(A+\bar{A}),
\eeq 
and requiring that 
\beq 
\frac{4\pi^3}{3\beta^3}(A+\bar{A})=\frac{\pi(c_{\mathrm{eff}}+\bar{c}_{\mathrm{eff}})}{6\beta^3},
\eeq 
we find
\beq 
A+\b A=\frac{c_{\mathrm{eff}}+\bar{c}_{\mathrm{eff}}}{8\pi^2}.
\eeq 
As argued earlier, we know that the function $c_{\mathrm{eff}}(s)$ defined in (\ref{defcef}) has the value $4\pi^2(f(r_{*})+\bar{f}(r_{*}))$ at conformal critical points characterized by a length scale $r_*$. The correlators above show that this values  is nothing but $\frac{c_{\mathrm{eff}}+\bar{c}_{\mathrm{eff}}}{2}$.  This completes our proof.

\medskip
\noindent {\em Remark.} The factor $8\pi^2$  arises once more from the fact that our normalization of the operators (\ref{operators}) is not the standard ``conformal normalization" that is used when defining the energy-momentum tensor. In the latter context one normally works with operators $\varepsilon:=-2\pi \tau $ and $\bar{\varepsilon}:=-2\pi \bar{\tau}$ and similarly for the hashed operators. Employing such operators we have then
that as $\beta\rightarrow \infty$ we  recover the results for CFT on the plane. In such a case the conformal correlators become:
 \beq 
\langle L_0| {\varepsilon}^{\real}(z) \varepsilon^\real(0)|R_0 \rangle=\frac{c_{\mathrm{eff}}}{2 z^4} \qquad \mathrm{and} \qquad \langle L_0| \bar{\varepsilon}^{\real}(\bar{z}) \bar{\varepsilon}^\real(0)|R_0 \rangle=\frac{\bar{c}_{\mathrm{eff}}}{2 \bar{z}^4}. 
\eeq 

\section{Testing the $c_{\rm{eff}}$-theorem: some examples}\label{sectest}

In order to illustrate the $c_\mathrm{eff}$-theorem, we now discuss a few interesting examples of RG flows where the conditions of the theorem are met, and some where they are not.

Firstly, we would like to discuss a QFT studied in \cite{fonseca} whose lagrangian density is
\beq
\mathcal{L}_{FZ}=\psi\bar\partial{\psi}+\bar{\psi}\partial \bar{\psi}+ im \bar{\psi}\psi + ih \sigma,
\label{fon}
\eeq
where $\psi$, $\bar{\psi}$ are the chiral components of the Majorana free Fermion field, $\sigma$ is the corresponding spin field and  $m,h\in \mathbb{R}$ with $m>0$. It was shown in \cite{fonseca} that the theory (\ref{fon}) displays an RG flow from the critical Ising to the critical Lee-Yang model, provided the ratio $\eta=m/|h|^{\frac{8}{15}}$  is fixed to a particular value. This describes the near-critical, universal region of a spin chain which was found earlier \cite{gehlen1}  to be in the Ising criticality class at one point, and in the Lee-Yang class along a curve emanating from that point defined by an algebraic relation between the two coupling constants involved (corresponding to $m$ and $h$ in \eqref{fon}). Both in the QFT \eqref{fon} and in the quantum chain, this critical curve separates a $\mathcal{PT}$-broken phase, where some energy eigenvalues occur in complex conjugated pairs, from an unbroken phase, where all eigenvalues are real. Therefore, at the critical curve and more generally in the $\mathcal{PT}$-unbroken phase, the theory (\ref{fon}) is the type of non-unitary model where we expect the $c_{\rm{eff}}$-theorem to hold. The explicit $\mathcal{PT}$-symmetry of the lagrangian in this case is:
\beq
\sigma \mapsto -\sigma,\quad \psi\mapsto i\psi,\quad \b\psi \mapsto i\b\psi,  \quad x\rightarrow -x,\quad i \rightarrow -i,
\eeq
which guarantees $\mathcal{PT}$-symmetry of the stress-energy tensor as per \eqref{PTass}. The flow \eqref{fon} on the critical curve satisfies the condition \eqref{order}, because $(c_\mathrm{eff})_{\rm UV}=0.5$ and $(c_\mathrm{eff})_{\rm IR}=0.4$, thus confirming $c_\mathrm{eff}$-theorem.

Further support for the existence of irreversible flows between non-unitary minimal models is the fact that, in some cases at least, it is possible to argue that they exhibit $\mathcal{PT}$-symmetry themselves. This may be shown by employing an effective Landau-Ginzburg description \cite{LGZ}. For the Lee-Yang minimal model \cite{Fisher}, the corresponding lagrangian density is
\beq\label{lyl}
\mathcal{L}_{\mathrm{LY}}=\frac{1}{2} (\partial_\mu \phi)^2+ im \phi^3
\eeq
A natural realization of $\mathcal{PT}$-symmetry in this case is the transformation 
\beq 
\phi \mapsto -\phi, \quad  i\mapsto -i \quad  \mathrm{and} \quad x\mapsto -x,
\label{ptre}
\eeq
under which the lagrangian is obviously invariant (this is a feature that is also well known from the study of the quantum-mechanical counterpart of this model \cite{BB}). 
More generally, it is known that unitary minimal models can be described by Landau-Ginzburg lagrangians with potentials which are even polynomials $V(\phi)=V(-\phi)$ with real coefficients \cite{LGZ}, where $\mathcal{PT}$-symmetry in the sense of (\ref{ptre}) is also present. Unfortunately there is no known generic Landau-Ginzburg description of the non-unitary minimal models, as discussed in detail in  \cite{amoroso}, even if the Lee-Yang case is well understood \cite{Fisher, LYCardy}. However, the presence of $\mathcal{PT}$-symmetry in all theories where the potential $V(\phi)$ either involves even powers of $\phi$ with real coefficients and/or odd powers of $\phi$ with imaginary coefficients, has been noted to hold quite generally, even in higher dimensions \cite{Codello}. Note that $\mathcal{PT}$-symmetry has been argued to be sufficient to ensure the stability of \eqref{lyl}, guaranteeing the spectrum to be real and bounded from below \cite{BBJ}.

Secondly, there are known RG flows where the condition (\ref{order}) is violated. Some of these examples have been discussed in \cite{nonmo1,nonmo2,nonmo25,nonmo3,nonmo4}. These examples would deserve more attention as they can only be reconciled with our result if some of the properties required for a $c_{\mathrm{eff}}$-theorem are not met. In  such theories  a decreasing monotonic function flowing between the UV and IR fixed points cannot exist. This could be explained in two possible ways: either $\mathcal{PT}$-symmetry is absent or broken, or the values of $c_{\rm{eff}}(s)$ at critical points do not coincide with the effective central charge as defined in (\ref{ceffcft}). The latter point is relevant because some of these theories have non-compact target space. This is associated with non-compact CFTs for which the result (\ref{Z}) is not always guaranteed to hold. However, in some cases it is hard to determine which of these two conditions is broken, in particular it is not easy to determine if $\mathcal{PT}$-symmetry is present or not.

An  example where the situation is simpler is the sine-Gordon model with purely imaginary coupling \cite{nonmo1, nonmo2, nonmo25}. This describes a non-unitary RG flow between two critical points, both with $c=1$. The  theory has lagrangian density of the form:
\beq \label{ish}
\mathcal{L}_{\mathrm{SG}}=\frac{1}{2} (\partial_\mu \phi)^2+ i\mu \cos\beta \phi,
\eeq 
where $\beta, \mu \in \mathbb{R}$ are coupling constants and $\phi$ is a scalar field. Using $p=\beta^2/(8\pi-\beta^2)$, it is natural to restrict to $p\geq 2$ \cite{nonmo1, nonmo2, nonmo25}.

If we employ the $\mathcal{PT}$ transformation (\ref{ptre}) it is clear that the cosine term in the lagrangian is not invariant. However, there are other possible realizations of $\mathcal{PT}$-symmetry, such as
\beq\label{ptsG}
\phi \mapsto \frc{\pi}\beta-\phi, \quad  i\mapsto -i \quad  \mathrm{and} \quad x\mapsto -x,
\eeq
which preserves the lagrangian. Thus, the theory possesses dynamical $\mathcal{PT}$-symmetry as per \eqref{PTass}, yet the strict inequalty \eqref{order} is violated and thus the $c_\mathrm{eff}$-theorem does not hold. We speculate that $\mathcal{PT}$-symmetry is in fact broken in this case: some of the energy eigenvalues are complex.

This speculation is based on the following two observations. First, in \cite{nonmo2,nonmo25}, it was observed that, when a TBA analysis of the massless scattering matrix proposed to describe this model is performed, the wrong UV value of the central charge is obtained. For $p> 3$ this was assumed to be due to technical difficulties, but seen to be more fundamental for $2\leq p \leq 3$. At least in this region, the real-energy massless states associated to this $S$-matrix might not form the complete set of states necessary for a TBA analysis. Indeed, in \cite{nonmo25}, it was conjectured that a certain pole of the scattering matrix, for $2\leq p < 3$, should be associated to a ``monstron" particle whose mass $M_m=e^{i\pi(3-p)/4}M$ (where $M$ is the ``intercept scale" of the massless spectrum) has positive imaginary part. This leads to exponentially growing amplitudes. The monstron particle provides energy eigenvalues with nonzero imaginary parts, thus breaking of $\mathcal{PT}$-symmetry and making the $c_\mathrm{eff}$-theorem inapplicable. Beyond this range of $p$, for instance for $3<p<7$, a simple analytic continuation suggests that the monstron's mass gets a negative imaginary part. This would give rise to decaying amplitudes which supposedly have a vanishing influence on the TBA analysis. However, one may speculate that it still breaks $\mathcal{PT}$-symmetry, and thus again makes the $c_\mathrm{eff}$-theorem inapplicable. In fact, by $\mathcal{PT}$-invariance of the hamiltonian, an ``anti-monstron" should also be present with complex conjugated mass. The transformation \eqref{ptsG} maps minima of the potential to maxima and {\em vice versa}. Thus, in a scattering theory whose particle spectrum is built with respect to a given supremum, only the monstron is visible; yet for $\mathcal{PT}$-symmetry to be applicable, one also needs to consider the other suprema, and thus the anti-monstron.

Second, quantum-group restrictions of the theory \eqref{ish} are known to correctly reproduce massless flows between unitary minimal models, lending support to the massless scattering matrix proposed. Something similar happens in the context of quantum-group invariant open XXZ chains, where boundary terms break hermiticity. These chains in general can be expected to possess states with complex eigenvalues, representing gain and loss processes. In this context, it is known that at ``roots of unity", diagonalizability does not hold and Jordan blocks appear (so that biorthogonality is broken), which quantum-group restrictions heal \cite{AB,PS,JK,Korff} giving rise to minimal models. A possible scenario is that such Jordan blocks arise as conjugate pairs collapse into a single real eigenvalue (such collapses are called ``exceptional points"), indeed suggesting the presence, at generic parameters, of eigenvalues with nonzero imaginary parts.

Finally, we note that there are known scaling functions, distinct from $c_\mathrm{eff}(s)$, which flow between two conformal critical points satisfying the property (\ref{order}) but which do so in a non-monotonic fashion. All the examples we are aware of arise in the context of the thermodynamic Bethe ansatz approach, where a natural scaling function $c^{\mathrm{TBA}}_\mathrm{eff}(s)$ can be defined which in unitary theories is known to encapsulate the same information as Zamolodchikov's $c$-function. Examples of such scaling functions have been presented in Fig.~1 and briefly discussed in the introduction. Examples of non-monotonic scaling functions have been presented in \cite{dorey:2000zb}. Since these scaling functions are different from our $c_{\rm{eff}}$-function and they satisfy (\ref{order}) it is clear that they do not provide counterexamples to our theorem. Obviously there are infinitely many continuous functions flowing between two points that may be constructed, both monotonic and non-monotonic. Our claim is that the function defined by (\ref{defcef}) exists and is monotonic under certain conditions. The existence of other non-monotonic functions does not challenge this claim.

\section{Conclusions and Outlook}\label{secconclu}

In this letter we have shown that a function with all the properties of a $c$-function may be constructed for non-unitary 1+1-dimensional QFTs under certain conditions, including crucially {\it unbroken $\mathcal{PT}$-symmetry} as expressed in \eqref{PTP}, \eqref{PTass}, and the ensuing positivity of the spectrum.  Thus, the requirement of unitarity, in the sense of a hermitian hamiltonian, is not necessary for the irreversibility of RG flows to hold. Besides unbroken $\mathcal{PT}$-symmetry, the properties we require (e.g. locality, Poincar\'e invariance etc.) are in fact very natural and routinely assumed to hold in QFT. Much of this paper is an effort to extricate some fundamental QFT concepts from unitarity and to show which properties are strictly necessary for the existence of a monotonic RG function and which are not.  The resulting scaling function is monotonically decreasing along RG flows and it is constant at critical points where, at least in rational models of CFT, it takes the value $\frac{c_{\mathrm{eff}}+\bar{c}_{\mathrm{eff}}}{2}$. In parity symmetric critical points this is the usual effective central charge introduced in \cite{ceff}.

There exist massive perturbations of non-unitary minimal models of CFT. In these cases, the $c_\mathrm{eff}$-theorem provides an alternative understanding of the positivity of $c_\mathrm{eff}$. It also guarantees that there cannot exist limit cycles with varying $c_\mathrm{eff}$ in non-unitary RG flows with unbroken $\mathcal{PT}$-symmetry.

The proof relies on somewhat abstract considerations of the non-hermitian stress-energy tensor components $\tau, \bar{\tau}$ and $\theta$ acting on a {\em bona fide} Hilbert space. For practical purposes, a very important open question is how to explicitly construct these operators and their ${\,}^\hash$ versions within the standard formulation of non-unitary CFTs (e.g. the non-unitary minimal series). In this standard formulation, the hamiltonian is hermitian, and one instead constructs the stress-energy tensor components in terms of generators of the Virasoro algebra of (possibly) negative central charge. These act on Verma modules, with an orthogonal basis (generated by the action of Virasoro operators on states created by primary fields) but with negative-norm states. Can we relate $\tau, \bar{\tau}$ and $\theta$ to such operators? Can we relate the states $|R_0\rangle$ and $\langle L_0|$ to the lowest-energy state $|\Delta\rangle, \langle \bar{\Delta}|$ of standard CFT formulations? These are questions we would like to address in the future.

We briefly discussed various examples where the assumptions and statement of the $c_\mathrm{eff}$-theorem can be checked, and an example where the statement does not hold, explaining how in this case $\mathcal{PT}$-symmetry appears to be broken. It would be important to provide more details on these examples, as well as other fully-worked cases, including a better understanding of the precise conditions of the $c_{\rm{eff}}$-theorem that are violated in those discussed in \cite{nonmo1,nonmo2,nonmo25,nonmo3,nonmo4}.

Another interesting question is how the entropic arguments used in \cite{Casinic} to prove Zamolodchikov's $c$-theorem can be generalized to give an alternative proof of the $c_{\rm{eff}}$-theorem.

\paragraph{Acknowledgement:} We would like to thank Patrick Dorey, Andreas Fring, Hubert Saleur, Germ\'an Sierra and Gerard Watts for useful discussions. We thank Hubert Saleur for bringing references \cite{nonmo2,nonmo3,nonmo4} to our attention and for sharing some private notes with us. We also thank Andreas Fring for bringing the special issue \cite{spePT}  and reference \cite{bi0} to our attention. 

Olalla Castro-Alvaredo and Benjamin Doyon are grateful to EPSRC for providing funding through the standard proposal ``Entanglement Measures, Twist Fields, and Partition Functions in Quantum Field Theory" under reference numbers EP/P006108/1 and EP/P006132/1. They also thank  the Physics Department of the University of Bologna, INFN and Elisa Ercolessi for hospitality and financial support during an extended visit in November 2016. 

Francesco Ravanini thanks INFN, in particular the Commission 4--Theory, for partial financial support through the grant GAST.

\begin{thebibliography}{10}

\bibitem{DMS}
P.~Di~Francesco, P.~Mathieu and D.~Senechal,
\newblock {Conformal Field Theory},
\newblock Springer  (1997).

\bibitem{BPZ}
A.~A. Belavin, A.~M. Polyakov and A.~B. Zamolodchikov,
\newblock {Infinite conformal symmetry in two-dimensional quantum field
  theory},
\newblock Nucl. Phys. {\bf B241}, 333--380 (1984).

\bibitem{zamc}
A.~B. Zamolodchikov,
\newblock Irreversibility of the flux of the renormalization group in a 2-D
  field theory,
\newblock JETP Lett. {\bf 43}, 730--732 (1986).

\bibitem{cth2}
D.~Boyanovsky and R.~Holman,
\newblock Zamolodchikov's $c$-theorem reexamined,
\newblock Phys. Rev. D {\bf 40}, 1964--1968 (1989).

\bibitem{Casinic}
H.~Casini and M.~Huerta,
\newblock A finite entanglement entropy and the $c$-theorem,
\newblock Phys. Lett.  {\bf B600}, 142--150 (2004).

\bibitem{roaming} Al.~B.~Zamolodchikov, Resonance factorized scattering and roaming trajectories, J. Phys. {\bf A39}, 12847-12862 (2006). 

\bibitem{CastroAlvaredo:2000ag}
O.~A. Castro-Alvaredo and A.~Fring,
\newblock {Renormalization group flow with unstable particles},
\newblock Phys. Rev. {\bf D63}, 021701 (2001).

\bibitem{BCN}
H.~Bl\"ote, J.~Cardy, and M.~Nightingale,
\newblock {Conformal invariance, the central charge, and universal finite size
  amplitudes at criticality},
\newblock Phys. Rev. Lett. {\bf 56}, 742--745 (1986).

\bibitem{Affleck}
I.~Affleck,
\newblock {Universal term in the free energy at a critical point and the
  conformal anomaly},
\newblock Phys. Rev. Lett. {\bf 56}, 746--748 (1986).

\bibitem{HolzheyLW94}
C.~Holzhey, F.~Larsen, and F.~Wilczek,
\newblock Geometric and renormalized entropy in conformal field theory,
\newblock Nucl. Phys. {\bf B424}, 443--467 (1994).

\bibitem{Latorre1}
G.~Vidal, J.~I. Latorre, E.~Rico, and A.~Kitaev,
\newblock Entanglement in quantum critical phenomena,
\newblock Phys. Rev. Lett. {\bf 90}, 227902 (2003).

\bibitem{Calabrese:2004eu}
P.~Calabrese and J.~L. Cardy,
\newblock Entanglement entropy and quantum field theory,
\newblock J. Stat. Mech. {\bf 0406}, P002 (2004).

\bibitem{ubi}
J.~{Cardy},
\newblock {The ubiquitous
`$c$': from the Stefan-Boltzmann law to quantum
  information},
\newblock J. Stat. Mech. {\bf 10},
  10004 (2010).

\bibitem{higherd1}
J.~L. Cardy,
\newblock {Is there a $c$ theorem in four-dimensions?},
\newblock Phys. Lett. {\bf B215}, 749--752 (1988).

\bibitem{atheorem}
Z.~Komargodski and A.~Schwimmer,
\newblock {On renormalization group flows in four dimensions},
\newblock JHEP {\bf 12}, 099 (2011).

\bibitem{Ftheorem}
R.~C. Myers and A.~Sinha,
\newblock Seeing a $c$-theorem with holography,
\newblock Phys. Rev.  {\bf D82}, 046006 (2010).

\bibitem{Ftheorem2}
H.~Casini and M.~Huerta,
\newblock Renormalization group running of the entanglement entropy of a
  circle,
\newblock Phys. Rev. {\bf D85}, 125016 (2012).

\bibitem{Casinia}
H.~Casini, E.~Teste, and G.~Torroba,
\newblock {The $a$-theorem and the Markov property of the CFT vacuum},
\newblock 1704.01870 (2017).

\bibitem{Fisher}
M.~Fisher,
\newblock Yang-Lee edge singularity and $\phi^3$ field theory,
\newblock Phys. Rev. Lett. {\bf 40}, 1610--1613 (1978).

\bibitem{LYCardy}
J.~L. Cardy,
\newblock Conformal invariance and the Yang-Lee edge singularity in two
  dimensions,
\newblock Phys. Rev. Lett. {\bf 54}, 1354--1356 (1985).

\bibitem{Cardymuss}
J.~Cardy and G.~Mussardo,
\newblock $S$-matrix of the Yang-Lee edge singularity in two-dimensions,
\newblock Phys. Lett. {\bf B225}, 275--278 (1989).

\bibitem{TCSA}
V.~P. Yurov and A.~B. Zamolodchikov,
\newblock {Truncated conformal space approach to scaling Lee-Yang model},
\newblock Int. J. Mod. Phys. {\bf A5}, 3221--3246 (1990).

\bibitem{Z}
A.~B. Zamolodchikov,
\newblock Two point correlation function in scaling Lee-Yang model,
\newblock Nucl. Phys. {\bf B348}, 619--641 (1991).

\bibitem{TW}
G.~Takacs and G.~Watts,
\newblock {Nonunitarity in quantum affine Toda theory and perturbed conformal
  field theory},
\newblock Nucl. Phys. {\bf B547}, 538--568 (1999).

\bibitem{TW2}
G.~Takacs and G.~Watts,
\newblock {RSOS revisited},
\newblock Nucl. Phys. {\bf B642}, 456--482 (2002).

\bibitem{gehlen1}
G.~von Gehlen,
\newblock {Critical and off critical conformal analysis of the Ising quantum
  chain in an imaginary field},
\newblock J. Phys. {\bf A24}, 5371--5400 (1991).

\bibitem{meandreas}
O.~Castro-Alvaredo and A.~Fring,
\newblock {A spin chain model with non-hermitian interaction: The Ising quantum
  spin chain in an imaginary field},
\newblock J. Phys. {\bf A42}, 465211 (2009).

\bibitem{fonseca} P.~Fonseca and A.~Zamolodchikov, Ising field theory in a magnetic field: analytic properties of the free energy, J. Stat. Phys. {\bf 110} 527--590 (2003).

\bibitem{ceff}
C.~Itzykson, H.~Saleur, and J.~Zuber,
\newblock {Conformal invariance of nonunitary two-dimensional models},
\newblock Europhys. Lett. {\bf 2}, 91 (1986).

\bibitem{BCDLR}
D.~Bianchini, O.~Castro-Alvaredo, B.~Doyon, E.~Levi, and F.~Ravanini,
\newblock {Entanglement entropy of non-unitary conformal field theory},
\newblock J.Phys. {\bf A48}, 04FT01 (2015).

\bibitem{BR} D.~Bianchini and F.~Ravanini, Entanglement entropy from corner transfer matrix in
                        Forrester Baxter non-unitary RSOS models, J. Phys. {\bf A49}, 154005 (2016). 


\bibitem{Saleur}
R.~Couvreur, J.~L. Jacobsen, and H.~Saleur,
\newblock {Entanglement in non-unitary quantum critical spin chains},
\newblock 1611.08506 (2016).


\bibitem{Lassig:1991an}
M.~L\"assig,
\newblock {New hierarchies of multicriticality in two-dimensional field
  theory},
\newblock Phys. Lett. {\bf B278}, 439--442 (1992).

\bibitem{Ahn:1992qi}
C.-R. Ahn,
\newblock {RG flows of nonunitary minimal CFTs},
\newblock Phys. Lett. {\bf B294}, 204--208 (1992).

\bibitem{TBA1}
A.~Zamolodchikov,
\newblock {Thermodynamic Bethe ansatz in relativistic models. Scaling three
  state Potts and Lee-Yang models},
\newblock Nucl. Phys. {\bf B342}, 695--720 (1990).

\bibitem{massless}
A.~Zamolodchikov,
\newblock {From tricritical Ising to critical Ising by thermodynamic Bethe
  ansatz},
\newblock Nucl. Phys. {\bf B358}, 524--546 (1991).

\bibitem{Martins:1991hi}
M.~J. Martins,
\newblock {The Thermodynamic Bethe ansatz for deformed $W A_{N-1}$ conformal field
  theories},
\newblock Phys. Lett. {\bf B277}, 301--305 (1992).

\bibitem{ravanini:1994pt}
F.~Ravanini, M.~Stanishkov, and R.~Tateo,
\newblock {Integrable perturbations of CFT with complex parameter: The $\mathcal{M}_{3,5}$
  model and its generalizations},
\newblock Int. J. Mod. Phys. {\bf A11}, 677--698 (1996).

\bibitem{dorey:2000zb}
P.~Dorey, C.~Dunning, and R.~Tateo,
\newblock {New families of flows between two-dimensional conformal field
  theories},
\newblock Nucl. Phys. {\bf B578}, 699--727 (2000).

\bibitem{nonmo1}
P.~Fendley, H.~Saleur, and Al.~B. Zamolodchikov,
\newblock {Massless flows I. The sine-Gordon and $O(n)$ models},
\newblock Int. J. Mod. Phys. {\bf A8}, 5717--5750 (1993).

\bibitem{nonmo2}
P.~Fendley, H.~Saleur, and Al.~B. Zamolodchikov,
\newblock {Massless flows II. The Exact $S$ matrix approach},
\newblock Int. J. Mod. Phys. {\bf A8}, 5751--5778 (1993).

\bibitem{nonmo25} Al.~B. Zamolodchikov, Thermodynamics of imaginary coupled sine-Gordon. Dense polymer finite-size scaling function, Phys. Lett. {\bf B335}, 436--443 (1994).

\bibitem{nonmo3}
N.~Read and H.~Saleur,
\newblock {Exact spectra of conformal supersymmetric nonlinear sigma models in
  two-dimensions},
\newblock Nucl. Phys. {\bf B613}, 409 (2001).

\bibitem{nonmo4}
J.~L. Jacobsen, N.~Read and H.~Saleur,
\newblock {Dense loops, supersymmetry, and Goldstone phases in two-dimensions},
\newblock Phys. Rev. Lett. {\bf 90}, 090601 (2003).

\bibitem{wigner}
E.~Wigner,
\newblock {Normal form of antiunitary operators},
\newblock J. Math. Phys. {\bf 1}, 409?413 (1960).

\bibitem{BB}
C.~M. Bender and S.~B\"ottcher,
\newblock {Real spectra in non-hermitian hamiltonians having $\mathcal{PT}$-symmetry},
\newblock Phys. Rev. Lett. {\bf 80}, 5243--5246 (1998).

\bibitem{Mo1}
A.~Mostafazadeh,
\newblock {Pseudo-hermiticity versus $\mathcal{PT}$-symmetry. The necessary condition for
  the reality of the spectrum},
\newblock J. Math. Phys. {\bf 43}, 205--214 (2002).

\bibitem{Mo2}
A.~Mostafazadeh,
\newblock {Pseudo-hermiticity versus $\mathcal{PT}$-symmetry 2. A Complete characterization
  of nonhermitian hamiltonians with a real spectrum},
\newblock J. Math. Phys. {\bf 43}, 2814--2816 (2002).

\bibitem{Mo3}
A.~Mostafazadeh,
\newblock {Pseudo-hermiticity versus $\mathcal{PT}$-symmetry 3: Equivalence of
  pseudohermiticity and the presence of antilinear symmetries},
\newblock J. Math. Phys. {\bf 43}, 3944--3951 (2002).

\bibitem{Carla}
C.~{Figueira de Morisson Faria} and A.~{Fring},
\newblock {Time evolution of non-hermitian hamiltonian systems},
\newblock Journal of Physics A Mathematical General {\bf 39}, 9269--9289 (2006).

\bibitem{Bender}
C.~M. Bender,
\newblock {Making sense of non-hermitian hamiltonians},
\newblock Rept. Prog. Phys. {\bf 70}, 947 (2007).

\bibitem{Mosta}
A.~Mostafazadeh,
\newblock {Pseudo-hermitian representation of quantum mechanics},
\newblock Int. J. Geom. Meth. Mod. Phys. {\bf 7}, 1191--1306 (2010).

\bibitem{spePT}
C.~Bender, A.~Fring, U.~G\"unther, and H.~Jones,
\newblock Quantum physics with non-hermitian operators,
\newblock J. Phys. {\bf A45}(44),
  440301 (2012).

\bibitem{bi0}
S.~Weigert,
\newblock $\mathcal{PT}$-symmetry and its spontaneous breakdown explained by
  anti-linearity,
\newblock J. Opt. B: Quantum Semiclassical Opt. {\bf 5}(3),
  S416 (2003).

\bibitem{bi1}
L.~N. Chang, Z.~Lewis, D.~Minic, and T.~Takeuchi,
\newblock {Biorthogonal quantum mechanics: super-quantum correlations and
  expectation values without definite probabilities},
\newblock J. Phys. {\bf A46}, 485306 (2013).

\bibitem{bi2}
D.~Brody,
\newblock Biorthogonal quantum mechanics,
\newblock J. Phys. {\bf A47}(3), 035305 (2014).

\bibitem{kubo}
R.~Kubo,
\newblock {Statistical-mechanical theory of irreversible processes. I. General
  theory and simple applications to magnetic and conduction problems},
\newblock J. Phys. Soc. Jpn. {\bf 12}, 570--586 (1957).

\bibitem{MS}
P.~C. Martin and J.~Schwinger,
\newblock {Theory of many-particle systems. I},
\newblock Phys. Rev. {\bf 115}, 1342 (1959).

\bibitem{Codello}
A.~Codello, M.~Safari, G.~P. Vacca, and O.~Zanusso,
\newblock {Leading CFT constraints on multi-critical models in $d> 2$},
\newblock JHEP {\bf 04}, 127 (2017).

\bibitem{amoroso} N.~Amoruso, Renormalization group flows between non-unitary conformal models, MSc Thesis, http://amslaurea.unibo.it/11308/, Universit\`a di Bologna (2015).

\bibitem{LGZ} A.~B.~Zamolodchikov, Conformal symmetry and multicritical points in two-dimensional quantum field theory, Sov. J. Nucl. Phys. {\bf 44}, 530 (1986).

\bibitem{BBJ} C.~M.~Bender, D. ~C.~Brody and H.~F.~Jones, Extension of $\mathcal{PT}$-symmetric quantum mechanics to quantum field theory with cubic interaction, 
Phys. Rev. {\bf D 70}, 025001 (2004); Erratum: Phys. Rev. {\bf D 71}, 049901 (2005).

\bibitem{AB} F.~C.~Alcaraz, M.~N.~Barber, M.~T.~Batchelor, R.~J.~Baxter and G.~R.~W.~Quispel, Numerical investigation of correlation functions for the 
$U_qSU(2)$ invariant spin-1/2 Heisenberg chain, J. Phys. {\bf A20}, 6397 (1987).

\bibitem{PS} V.~Pasquier and H.~Saleur, Common structures between finite systems and conformal field theories through quantum groups, Nucl. Phys. {\bf 330}, 523--556 (1990).

\bibitem{JK} G. J\"uttner and M.~Karowski, Completeness of ``good'' Bethe ansatz solutions of a quantum group invariant Heisenberg model Nucl. Phys. {\bf B430}, 615 (1994).

\bibitem{Korff} C.~Korff and R.~A.~Weston, PT symmetry on the lattice: the quantum group Invariant XXZ spin-chain, J. Phys. {\bf A40}, 8845 (2007).
\end{thebibliography}

\end{document}